\documentclass[fleqn,usenatbib]{mnras}

\usepackage{newtxtext,newtxmath}

\usepackage[T1]{fontenc}

\DeclareRobustCommand{\VAN}[3]{#2}
\let\VANthebibliography\thebibliography
\def\thebibliography{\DeclareRobustCommand{\VAN}[3]{##3}\VANthebibliography}

\usepackage{graphicx}	
\usepackage{amsmath}	

\newcommand\totalGALaligned{4522\,}
\newcommand\totalGALmisaligned{247\,}

\newcommand\mergerGALaligned{1710\,}
\newcommand\mergerGALmisaligned{70\,}
\newcommand\totalgal{4769\,}
\newcommand\totalmerger{1780\,}
\newcommand\totalAGN{274\,}

\newcommand\AGNfracnonmerger{5.1\%\,}
\newcommand\AGNfracmerger{6.8\%\,}

\newcommand\AGNfrac{5.7\%\,}

\newcommand\totalfracAGNaligned{5.2$^{+0.4}_{-0.3}\%$\,}
\newcommand\totalfracAGNmisaligned{15$^{+2}_{-2}\%$\,}
\newcommand\mergerfracAGNaligned{6.2$^{+0.6}_{-0.5}\%$\,}
\newcommand\mergerfracAGNmisaligned{20$^{+6}_{-4}\%$\,}

\newcommand\totalGALearly{13\%\,}
\newcommand\totalGALlate{87\%\,}

\newcommand\totalGALmergerlate{91\%\,}
\newcommand\totalGALmergerearly{9\%\,}

\newcommand\totalGALalignedlate{90\%\,}
\newcommand\totalGALalignedearly{10\%\,}
\newcommand\totalGALmisalignedlate{38\%\,}
\newcommand\totalGALmisalignedearly{62\%\,}

\newcommand\LateInAGN{81\%\,}
\newcommand\EarlyInAGN{19\%\,}

\newcommand\AGNlateEitherDeltaPA{5\%\,}
\newcommand\AGNearlyEitherDeltaPA{7\%\,}

\title[Black hole activity in interacting galaxies]{Kinematic misalignment as a driver of black hole activity in galaxies with external interactions}

\author[S. I. Raimundo et al.]{Sandra I. Raimundo $^{1,2}$\thanks{E-mail: s.raimundo@soton.ac.uk}, Rogerio Riffel$^{3}$, Song-lin Li$^{4,5}$,   Cristina Ramos Almeida$^{6,7}$, Sandro Rembold$^{8}$ \newauthor Rogemar A. Riffel$^{8}$, Thaisa Storchi-Bergmann$^{3}$, Marianne Vestergaard$^{2,9}$, José L. Tous$^{1}$
\\
$^{1}$Physics and Astronomy, University of Southampton, Highfield, Southampton, SO17 1BJ, UK\\
$^{2}$DARK, Niels Bohr Institute, University of Copenhagen, Jagtvej 155, Copenhagen N, 2200, Denmark\\
$^{3}$ Departamento de Astronomia, Instituto de F\'\i sica, Universidade Federal do Rio Grande do Sul, CP 15051, 91501-970, Porto Alegre, RS, Brazil \\
$^{4}$Research School of Astronomy and Astrophysics, Australian National University, Canberra, ACT 2611, Australia \\
$^{5}$ARC Centre of Excellence for All-Sky Astrophysics in 3 Dimensions (ASTRO 3D), Canberra, ACT 2611, Australia \\
$^{6}$ Instituto de Astrofísica de Canarias, Calle Vía Láctea, s/n, 38205, La Laguna, Tenerife, Spain\\
$^{7}$ Departamento de Astrofísica, Universidad de La Laguna, 38206, La Laguna, Tenerife, Spain\\
$^{8}$ Departamento de Física, Centro de Ciências Naturais e Exatas, Universidade Federal de Santa Maria, 97105-900, Santa Maria, RS, Brazil\\ 
$^{9}$ Steward Observatory, University of Arizona, Tucson, AZ, USA
}

\date{Accepted XXX. Received YYY; in original form ZZZ}

\pubyear{2024}

\begin{document}
\label{firstpage}
\pagerange{\pageref{firstpage}--\pageref{lastpage}}
\maketitle

\begin{abstract}
The process of Active Galactic Nuclei (AGN) fuelling relies on the transport of gas across several orders of magnitude in physical scale until the gas reaches the supermassive black hole at the centre of a galaxy. This work explores the role of kinematically misaligned gas in the fuelling of AGN in a sample of \totalgal\ local galaxies from the MaNGA survey. We investigate for the first time the relative role of external interactions and the presence of kinematic misalignment as mechanisms to explain the observed increase in AGN fraction in galaxies with large stellar to gas kinematic misalignment ($>$45$^{\circ}$). Using a sample of galaxies with evidence of recent external interactions we find that there is a significantly higher fraction of AGN in those where a large stellar to gas kinematic misalignment is observed (\mergerfracAGNmisaligned) compared with \mergerfracAGNaligned in galaxies where no kinematic misalignment is observed. We determine that gas to stellar misalignment has an important role in the fraction of AGN observed, increasing the AGN fraction beyond the potential effect of external interactions. This result demonstrates the importance of misaligned structures to the fuelling of supermassive black holes.

\end{abstract}




\section{Introduction}
The process of supermassive black hole growth via gas accretion requires gas to be transported from galaxy scales to the accretion disc ($\sim$10$^{16}$ cm, e.g. \citealt{morgan10}). The fuelling gas may originate from the host galaxy or from its immediate external environment. In both cases, the gas needs to lose angular momentum to be transported to the vicinity of the supermassive black hole, requiring dynamical mechanisms that are able to cause this angular momentum loss, such as galaxy interactions or large-scale bars, amongst other processes (e.g. see \citealt{shlosman90}, \citealt{martini04}, \citealt{storchi-bergmann19} and \citealt{combes23} for reviews). This process of black hole gas fuelling is inherently connected to the powering of active galactic nuclei (AGN) and plays a decisive role in our understanding of AGN triggering mechanisms and the overall luminosity of AGN.

Although direct correlations between fuelling mechanisms and AGN are difficult to find (\citealt{martini03}), several candidate mechanisms and observational evidence to support them, have been put forward to explain the fuelling of AGN at different luminosity and redshift ranges, e.g. galaxy mergers (e.g. \citealt{ramos-almeida11}, \citealt{ramos-almeida12}, \citealt{fischer15}, \citealt{koss18}, \citealt{pierce23}, \citealt{araujo23}, \citealt{comerford24}), bars, nuclear spirals and gravitational torques (e.g. \citealt{shlosman89}, \citealt{garcia-burillo05}, \citealt{schnorr-muller11}, \citealt{schnorr-muller14}, \citealt{kim17}, \citealt{delmoral-castro20}, \citealt{audibert21}, \citealt{rembold24}), stellar mass loss or interstellar medium turbulence (e.g. \citealt{davies07}, \citealt{choi24}, \citealt{riffel24}). In this work we focus on an AGN fuelling mechanism that has recently been discovered observationally, namely that of stellar to gas kinematic misalignment \citep{raimundo23}. In galaxies dominated by secular evolution, it is expected that gas and stars share a similar axis of rotation. That is in general not the case for galaxies that undergo dynamical interactions causing external gas accretion. If the amount of accreted material is significant compared to the aligned gas and stars already present, the accreted material may end up with an angular momentum vector orientation that is distinct from that of the main stellar body of the galaxy (\citealt{haynes84}, \citealt{bertola92}, \citealt{sancisi08}). This can also be the consequence of kinematic misalignment of the halo itself, as has been shown by numerical simulations \citep{lagos15}. Notable examples of extreme angular momentum differences are polar ring / polar disc galaxies (\citealt{sersic67}, \citealt{schweizer83}, \citealt{bournaud&combes03}) or galaxies with counter-rotating cores/discs (e.g. \citealt{franx&illingworth88}, \citealt{rubin92}, \citealt{kannapan&fabricant01}, \citealt{pizzella04}, \citealt{sil'chenko09}, \citealt{raimundo13}, \citealt{bevacqua22}, \citealt{katkov24}). Large misalignment angles between stellar and gas motions can therefore be used to identify candidate galaxies that went through a past external accretion event, such as a major or minor merger, a galaxy flyby or gas infall from a neighbour galaxy (\citealt{bertola92}, \citealt{davis&bureau16}, \citealt{li21}) and to investigate the timescale and consequence of these processes (e.g. \citealt{davis11}, \citealt{ilha19}, \citealt{raimundo21}, \citealt{khoperskov21}, \citealt{ristea22}, \citealt{roier22}, \citealt{xu22}, \citealt{cenci24}, \citealt{baker25}). 

Observations point towards a connection between stellar to gas misalignment and a higher fraction of AGN activity. \cite{raimundo23} have shown that galaxies with large kinematic misalignment angles ($>$45$^{\circ}$) have a higher fraction of AGN than galaxies without misalignment, suggesting that stellar to gas kinematic misalignment and/or the external accretion event that originated it, is connected with the fuelling of AGN. Black hole fuelling requires gas and dynamical mechanisms to promote the loss of angular momentum, and both these conditions can be met in galaxies with large misalignment. Firstly, external accretion events provide a supply of gas to the host galaxy, which is potentially significant for early type galaxies without a large reservoir of native gas (\citealt{simoes-lopes07}, \citealt{davies14}, \citealt{raimundo17}, \citealt{khim20}). Secondly, it has been shown from simulations that misaligned structures (stellar/stellar or stellar/gas misalignment) promote the loss of angular momentum and the flow of gas towards the centre of the galaxy, via for example stellar torques, shocks in the gas or dynamical friction (\citealt{thakar&ryden96},  \citealt{negri14}, \citealt{vandevoort15}, \citealt{capelo&dotti17}, \citealt{taylor18}, \citealt{starkenburg19}, \citealt{duckworth20},\citealt{khoperskov21}), with recent observational evidence supporting that hypothesis (\citealt{raimundo21}, \citealt{zhou22}, \citealt{raimundo23}).

Since kinematic misalignment is very often a consequence of an external accretion event, the two factors (external gas supply and kinematic misalignment) can be linked when investigating their effect on AGN. While they may both contribute to the observed increase in AGN \citep{raimundo23}, it is unclear if the isolated effect of kinematic misalignment makes a significant contribution to increase the fraction of AGN. This is particularly important since mergers of galaxies, which cause $\sim 10-20$\% of misalignments in massive galaxies (M$_{*}>10^{10}$M$_{\odot}$) - \citealt{baker25}, have been observed to be connected with higher fractions of AGN (e.g. \citealt{rembold24}, \citealt{comerford24} for the MaNGA sample).
To answer this question, we investigate whether the presence of misalignment has a prevalent effect in the increase of AGN fraction in galaxies with both signatures of external interactions and stellar to gas misalignment. In Section~\ref{sec:data_analysis} we describe the data and the methods used to determine kinematic position angles, to identify AGN and to identify galaxies with recent interactions. In Section~\ref{sec:results_discussion} we show how the AGN fraction varies as a function of misalignment angle, and discuss the driving mechanism for the observed increased fraction of AGN in galaxies with strong kinematic misalignment.

\section{Data analysis}
\label{sec:data_analysis}
We used observations from the Data Release 17 of the Mapping Nearby Galaxies
at Apache Point Observatory (MaNGA)  optical integral field unit survey of $\sim$ 10 000 local galaxies (\citealt{bundy15}, \citealt{abdurro'uf22}). The survey covers a redshift range of 0.01 $<$ z $<$ 0.15 and stellar masses M$_{\star} > 10^{9} $M$_{\odot}$. In this work we use the MaNGA 2D maps of gas and stellar properties as analysed and compiled in the form of MEGACUBES\footnote{manga.linea.org.br} and presented in detail by \cite{riffel23}.

\subsection{Measurement of kinematic position angles}
The global kinematic position angles represent the mean motion of the stars (PA$_{\rm stellar}$) and of the gas (PA$_{\rm gas}$), and are determined from two-dimensional maps of stellar and gas velocity, respectively. The position angle is defined as the angle between the north and the line that
connects the absolute maxima of the velocity. To calculate the position angles we follow the same approach as in \cite{raimundo23}, applying the \textsc{fit\_kinematic\_pa} algorithm \citep{krajnovic06}  to the maps of ionised gas velocity (obtained from measuring the velocity of the H$\alpha$ emission lines) and stellar velocity from the MEGACUBES distribution \citep{riffel23}.

\begin{figure}
\includegraphics[width=\columnwidth]{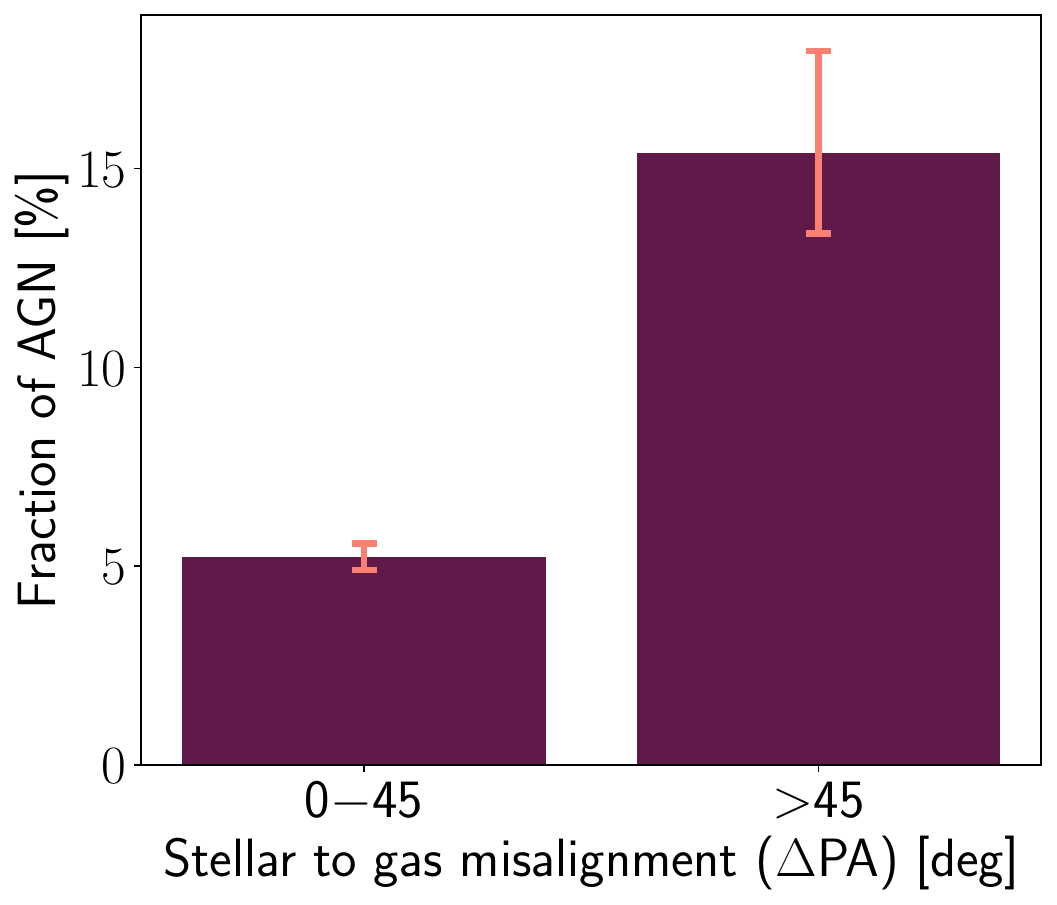}
\caption{Fraction of AGN as a function of stellar to gas kinematic misalignment ($\Delta$PA in degrees) for the galaxies in the MaNGA survey. Galaxies with significant kinematic misalignment ($\Delta$PA $\ge$ 45$^{\circ}$) show a higher fraction of AGN: \totalfracAGNmisaligned compared with \totalfracAGNaligned for the galaxies with $\Delta$PA $<$ 45$^{\circ}$. The error bars indicate the 68$\%$ confidence intervals calculated using the beta distribution quantile technique for a binomial population \citep{cameron11}.}
\label{fig:manga_aligned_misaligned}
\end{figure}

\begin{figure}
\includegraphics[width=\columnwidth]{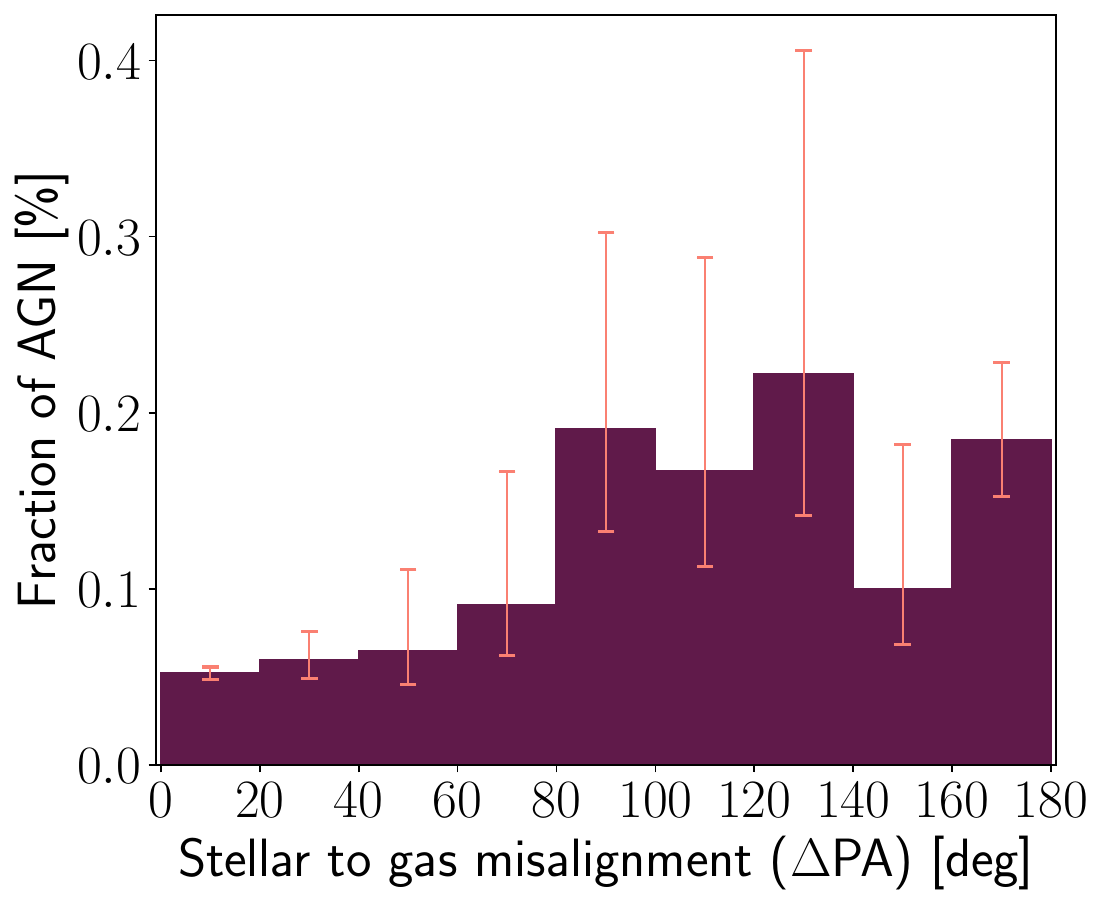}
\caption{Fraction of AGN as a function of stellar to gas kinematic misalignment ($\Delta$PA in degrees) for the galaxies in the MaNGA survey, divided into 20$^{\circ}$ bins. This figure is similar to Fig.~\ref{fig:manga_aligned_misaligned} but with smaller number statistics due to the reduced number of galaxies per bin. The error bars indicate the 68$\%$ confidence intervals calculated using the beta distribution quantile technique for a binomial population \citep{cameron11}.}
\label{fig:manga_aligned_misaligned_20deg}
\end{figure}

In this work we are interested in the difference between the direction of motion of stars and gas in a galaxy as projected in the sky plane, i.e. how misaligned the stellar and gas rotation axes are. The misalignment angle ($\Delta$PA) for each galaxy is calculated from the difference between the global stellar and gas kinematic angles: $\Delta$PA = $|$PA$_{\rm stellar}$ - PA$_{\rm gas}|$ and can vary between 0$^{\circ}$ (perfect alignment) and 180$^{\circ}$ (counter-rotation). We would like to highlight that these angles reflect global gas and stellar motions measured at scales of hundreds of parsecs to kiloparsecs, and not local small scale variations. To determine $\Delta$PA we require that both PA$_{\rm stellar}$ and PA$_{\rm gas}$ can be measured accurately. Our quality cuts are similar to those implemented by \cite{raimundo23}: we only use velocity map pixels where the uncertainties in the velocities are lower than 30 km/s, and require the uncertainty on the measured $\Delta$PA to be lower that 30$^{\circ}$. We also require a minimum of 20 good pixels to determine the stellar or gas PA. In total we have a sample of \totalgal galaxies for which the misalignment angle was determined. This is our parent sample. Within this sample, \totalGALlate are late-type galaxies and \totalGALearly are early-type galaxies, according to the visual morphology classification of  \cite{vazquez-mata22}.

\subsection{AGN identification}
To identify AGN we combine two different approaches. The first is the AGN optical identification based on the spatially-resolved Baldwin, Phillips \& Terlevich \citep{baldwin81} diagrams as described by \cite{raimundo23}.
The second is the AGN identification by \cite{comerford24} using multiwavelength observations and catalogues. The AGN classification of \cite{raimundo23} uses spatially resolved narrow emission line ratios to identify AGN, and more details can be found in that work. The \cite{comerford24} AGN identification includes broad emission line AGN detected in the Sloan Digital Sky Survey (SDSS), AGN classified in infrared (Wide-Field Infrared Survey Explorer - WISE), radio (Faint Images of the Radio Sky at Twenty cm - FIRST) and X-ray (Swift Burst Alert Telescope - Swift/BAT) catalogues. We combine these two samples of AGN and match them to the sample of \totalgal galaxies for which the misalignment angle was determined. Our final sample of AGN with measured $\Delta$PA contains a total of \totalAGN AGN (out of N=\totalgal galaxies with measured kinematic angles, i.e. an AGN fraction of \AGNfrac). Among the AGN sample, \LateInAGN of them are in late-type galaxies while \EarlyInAGN are in early-type galaxies. The stellar mass range of the AGN host galaxies is distributed from M$_{\star}\sim$10$^{9}$ - 10$^{11}$ M$_{\odot}$ with a peak at $\sim$10$^{10.4}$M$_{\odot}$. The AGN [O III] luminosity range is typically L$_{\rm OIII}\sim$10$^{39}$ - 10$^{41.5}$ erg/s, which indicates that these are low to moderate luminosity AGN \citep{winiarska25}. The low luminosity range of the AGN supports previous findings \citep{raimundo23}, that the gas kinematic misalignment observed in AGN hosts is not driven by potential AGN outflows (\citealt{ilha19}, \citealt{khoperskov21}, \citealt{raimundo23}). This is because of the observed trend between outflow size and AGN luminosity (e.g. \citealt{kim23}). In addition to the AGN selection outlined above, we also used, as a comparison, the AGN identification presented by \cite{rembold17} and \cite{riffel23} based on the diagnostic diagrams of  Baldwin, Phillips \& Terlevich \citep{baldwin81} and that of the equivalent width of H$\alpha$ versus [N II]/H$\alpha$ (WHAN) of \citep{cidfernandes10}. The results will be discussed in Section \ref{sec:results_discussion}.

\subsection{Signatures of galaxy interactions}
To identify galaxies that underwent recent interactions in the MaNGA sample, we combine two methods: the visual classification of \cite{li21} for MaNGA galaxies that show evidence of a past interaction, and the machine learning identification of major and minor mergers from \cite{nevin23} as used by \cite{comerford24} in their analysis of MaNGA. \cite{li21} used deep images from the Dark Energy Spectroscopic Instrument (DESI) Legacy Imaging Surveys \citep{dey19} to visually search for evidence of past interactions in the MaNGA sample. We use their sample of 538 galaxies with merging features or evidence of strong interaction with companions identified in the MaNGA sample, such as tidal features, distortions/asymmetries or shells. From the sample of \cite{comerford24} we use the galaxies that have a probability above 50\% of being either major mergers or minor mergers. Out of our initial sample of \totalgal MaNGA galaxies with well measured kinematic angles, \totalmerger galaxies show evidence of interactions according to either one of the methods described above (\citealt{li21} or \citealt{comerford24}). Out of \totalmerger galaxies with interactions, 1662 are selected using the \citealt{comerford24} method. An extra 132 galaxies are identified in the \citealt{li21} catalogue only. Within the sample of \totalmerger galaxies, \totalGALmergerlate are late-type galaxies and \totalGALmergerearly are early-type galaxies, according to the visual morphology classification of \cite{vazquez-mata22}. Since the identification of interactions relies on photometric features, it means that we are identifying relatively recent interactions, before the features evolved and became too faint to be identified.

\begin{figure}
\includegraphics[width=\columnwidth]{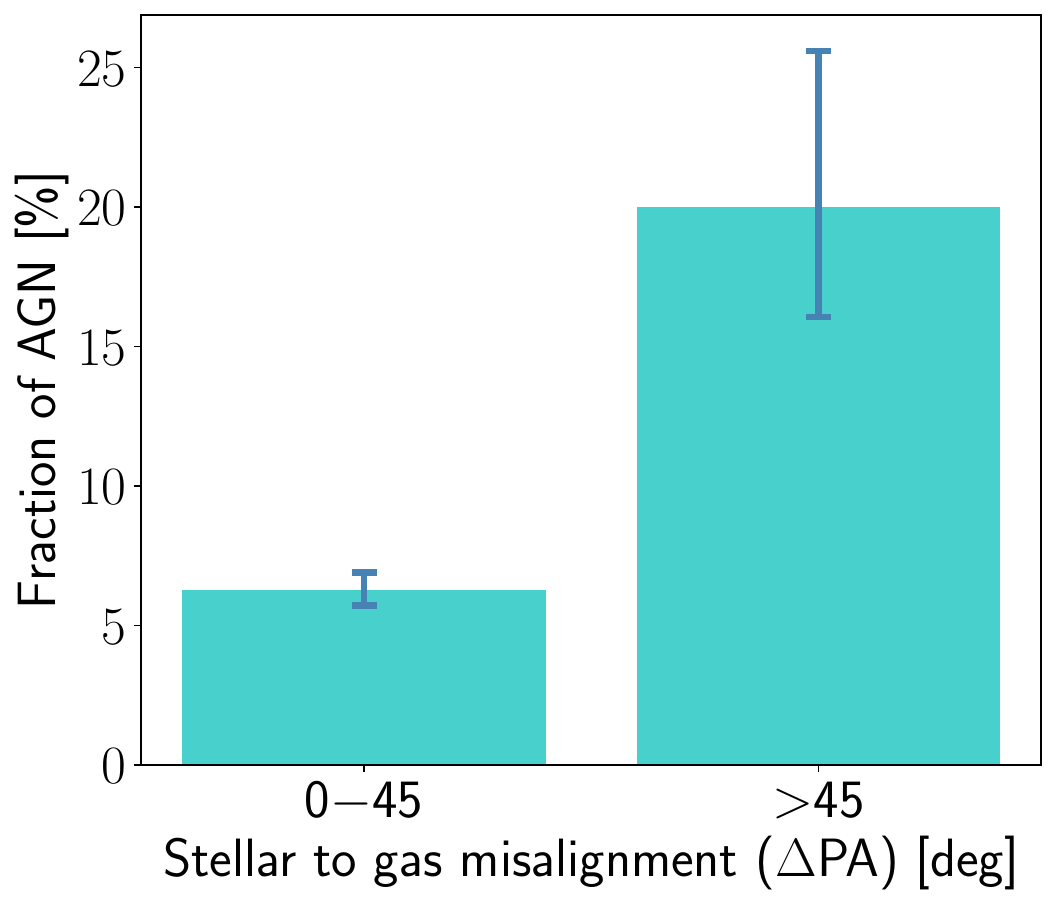}
\caption{Fraction of AGN as a function of stellar to gas kinematic misalignment ($\Delta$PA in degrees) for the galaxies in the MaNGA survey that show signatures of a past interaction. Galaxies with significant kinematic misalignment ($\Delta$PA $\ge$ 45$^{\circ}$) show a higher fraction of AGN, indicating that misalignment is a driving mechanism for the increased fraction of AGN and not only due to a dynamical interaction, such as a merger. The relative fractions are \mergerfracAGNmisaligned for galaxies with $\Delta$PA $\ge$ 45$^{\circ}$ compared to  \mergerfracAGNaligned for the galaxies with $\Delta$PA $<$ 45$^{\circ}$.The error bars indicate the 68$\%$ confidence intervals calculated using the beta distribution quantile technique for a binomial population \citep{cameron11}.}
\label{fig:manga_mergers}
\end{figure}

\section{Results and Discussion}
\label{sec:results_discussion}
In the following sub-sections we investigate the trend between AGN fraction and misalignment angle for the MaNGA sample and study if the same trend is found in the sub-sample of \totalmerger MaNGA galaxies that have signatures of past interactions. 

\subsection{Fraction of AGN in galaxies with misalignment}
To investigate if the fraction of AGN changes for galaxies with and without a significant misalignment between stellar and gas kinematics, we follow the method of \cite{raimundo23}. We start with the full sample of MaNGA galaxies for which the measurement of stellar and gas kinematic angles was possible (N=\totalgal galaxies). We then divide the sample into two groups: those with $\Delta$PA$<$45$^{\circ}$, which are galaxies for which stellar and gas kinematic angles are likely aligned within the uncertainties (N=\totalGALaligned), and those with $\Delta$PA$\ge$45$^{\circ}$, which are galaxies for which a significant misalignment is observed (N=\totalGALmisaligned). In Fig.~\ref{fig:manga_aligned_misaligned} we show the fraction of AGN in these two misalignment bins. We find an AGN fraction of \totalfracAGNaligned for galaxies with $ 0^{\circ} \leq \Delta$PA $< 45^{\circ}$ and a fraction of \totalfracAGNmisaligned for galaxies with $\Delta$PA $\ge45^{\circ}$. It is clear that there is a significantly higher ($>$3$\sigma$ level)  fraction of AGN in galaxies with misaligned gas and stars for the MaNGA sample, similar to what is found for the SAMI survey, which covers a similar redshift range but has a smaller galaxy sample size \citep{raimundo23}. We also observe an increased fraction of AGN if we consider polar galaxies ($\Delta$PA$\sim$90$^{\circ}$) or counter-rotating galaxies ($\Delta$PA$\sim$180$^{\circ}$) separately. In Fig.~\ref{fig:manga_aligned_misaligned_20deg} we show the same sample but divided into 20$^{\circ}$ bins. While there are fewer galaxies per bin, this figure is useful to illustrate that the main differences in AGN fraction occurs for $\Delta$PA $\gtrsim$ 60$^{\circ}$.

\begin{figure*}
\includegraphics[width=15cm]{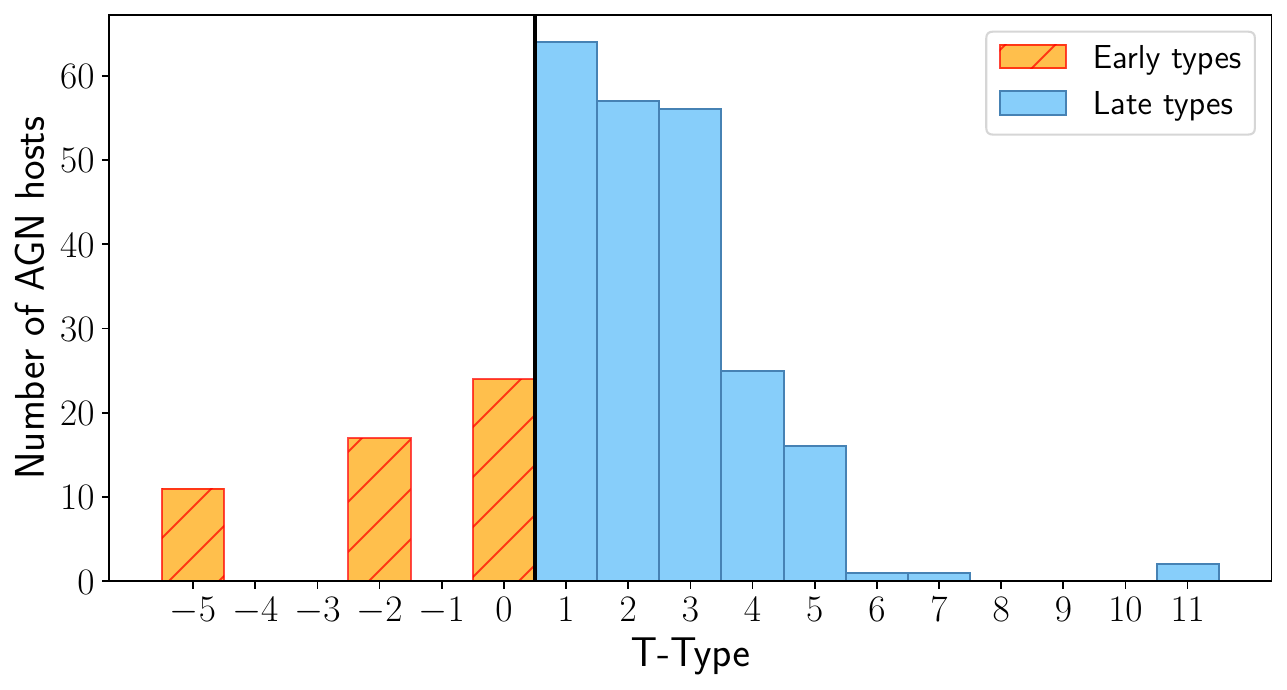}
\caption{Distribution of AGN host morphology for the \totalAGN AGN in our sample. The morphology is given by the T-Type morphology indicator which takes on discrete integer values \citep{vazquez-mata22} . Each bin is centered at the specific discrete value in T-Type with a width of 0.5 for visualisation purposes. The orange hatched histogram shows the AGN host galaxies that are early-type galaxies (T-Type < 1), while the blue histogram shows the AGN host galaxies that are late types (T-Type $\ge$ 1).}
\label{fig:host_morph}
\end{figure*}

It is known that early-type galaxies have a higher fraction of kinematic misalignment than late types (e.g. \citealt{bertola91}, \citealt{pizzella04}, \citealt{davis11}, \citealt{raimundo23}). To check that the trend that we observe is not driven by a preference of AGN for early-type hosts, we use the T-Type \citep{devaucouleurs59} visual morphology classifications as presented in \cite{vazquez-mata22} to investigate the distribution of AGN host morphologies.  The T-Type values can be related to the Hubble classification in that T-Type values < 1 correspond to early-type galaxies and T-Type values $\ge$ 1 to late-type galaxies. We show our results in Fig.~\ref{fig:host_morph}, with a histogram of AGN host numbers as a function of T-Type. The orange histogram shows early-type hosts, while the blue histogram shows late-type hosts. Within the sample of MaNGA AGN galaxies with measured stellar and gas kinematic angles, most hosts are late types. We also carry out the quantitative exercise of estimating the difference in AGN fraction between each bin: $\Delta$PA $< 45^{\circ}$ and $\Delta$PA $\ge45^{\circ}$) that would be expected from having the combined effect of a higher fraction of AGN in early-type galaxies and a higher fraction of early-type galaxies with misalignment. Within the \totalGALaligned aligned galaxies of the parent sample, \totalGALalignedlate are late-type and \totalGALalignedearly early-type. Within the \totalGALmisaligned misaligned galaxies there are \totalGALmisalignedlate late-type and \totalGALmisalignedearly early-types. In the overall MaNGA sample with or without measured kinematic angles the fraction of AGN in late-types is \AGNlateEitherDeltaPA and in early-types \AGNearlyEitherDeltaPA. Based on these values and without any additional trend, the expected fraction of AGN in $\Delta$PA $<$ 45$^{\circ}$ would be $\sim$ 5\% and in $\Delta$PA $\geq$ 45$^{\circ}$ would be $\sim$ 6\%, resulting in a difference of $\sim$ 1\%, which is a much smaller difference than what we see in our sample ($\sim$9-10\% - Fig.~\ref{fig:manga_aligned_misaligned}).
Fig.~\ref{fig:host_morph} and the quantitative exercise above support the argument that the trend for a higher fraction of AGN in misaligned galaxies is not due to a double correlation between AGN and early-type morphology and early-type morphology and misalignment but needs an additional effect which our work argues is the presence of misaligned gas. 

\subsection{Driving mechanism - galaxy interactions or misaligned gas?}
In this work we want to separate the effect of external interactions from kinematic misalignment in driving the increase in AGN fraction. Kinematic misalignment is the result of external accretion but not all external accretion and interactions will result in a large kinematic misalignment. Depending on the configuration, the external inflow of material may end up in kinematic alignment ($\Delta$PA $= [0^{\circ} - 45^{\circ}]$) or relaxing into kinematic alignment with a particular time delay after the interaction. For example polar disc galaxies are found to settle into co/counter-rotation within 1-3 Gyr \citealt{khoperskov21}. Kinematic misalignment can also be longer lived than the time during which morphological signatures of interactions are observed (\citealt{starkenburg19}, \citealt{ebrova21}). Therefore, galaxies may show kinematic misalignment but no signatures of the interaction that generated it. To determine which of the mechanisms is dominant in increasing the fraction of active black holes, we select the sample of MaNGA galaxies with signatures of recent interactions as our parent sample, finding a total of \totalmerger galaxies (\mergerGALaligned with $0^{\circ} \leq \Delta$PA $< 45^{\circ}$ and \mergerGALmisaligned with $\Delta$PA $\ge45^{\circ}$). These galaxies may have had different types of external interactions, therefore in this work we do not establish the role of a particular external interaction but analyse all types of external interactions as a whole. We then determine the fraction of AGN on the sub-samples of galaxies with aligned and misaligned stellar to gas kinematics. If external accretion (through some form of interaction or merger activity) was the most important factor and gas kinematical misalignment is irrelevant, then we should expect to see no statistical difference in the $\Delta$PA distribution for the galaxies with an AGN (for this galaxy subset with evidence of interactions). In Fig.~\ref{fig:manga_mergers} we show the fraction of AGN as a function of kinematic misalignment angle. We find an AGN fraction of \mergerfracAGNaligned for galaxies with $ 0^{\circ} \leq \Delta$PA $< 45^{\circ}$ and a fraction of \mergerfracAGNmisaligned for galaxies with $\Delta$PA $\ge45^{\circ}$, different by $\sim$3$\sigma$. A difference is also found between the two populations (albeit with smaller number statistics) if we use only the \cite{riffel23} sample of AGN (which includes a WHAN diagnostic - \citealt{cidfernandes10}) or only the \citealt{comerford24} AGN sample. With our main sample shown in Fig.~\ref{fig:manga_mergers}, it is clear that even within the sample of galaxies with interactions (which includes major/minor mergers and different stages of interaction), there is a significantly higher fraction of AGN in galaxies with misaligned gas and stars ($\Delta$PA $>$ 45$^{\circ}$) than for kinematically aligned gas and stars. This shows that the presence of misalignment is connected with an additional increase in the fraction of AGN, being a potentially important ingredient for the activation of the black hole.

\cite{comerford24} recently found that among galaxies with signatures of past major or minor mergers in MaNGA, there is an increased number of AGN.
Comparing Figs.~\ref{fig:manga_aligned_misaligned} and \ref{fig:manga_mergers}, we find a tentative increase in the overall fraction of AGN in all galaxies (aligned and misaligned combined) from \AGNfracnonmerger in the initial sample to \AGNfracmerger in the sample of galaxies with signatures of past interactions. Both samples consist of galaxies for which stellar and gas kinematic angles can be determined, and therefore neither is an unbiased sample. In any case, the increased fractions we observe is in line with the results of \cite{comerford24}.

We test whether the trend we observe in Fig.~\ref{fig:manga_mergers} is also present if we split the sample by stellar mass or morphology. The main limitation we have is the small number statistics, which affect the significance of the results. In any case, we find a difference of $>2.5\sigma$ between the two populations even if we split the sample into low and high stellar masses (using a threshold of log($M_{*}/M_{\odot}$) = 9.5, 10 or 10.5), showing that the trend we observe is not driven by differences in stellar mass. We also find evidence that the trend is still present when the sample is split into spirals versus ellipticals and S0s, suggesting that the trend we see is not driven by morphology.

Previous work has shown a connection between the presence of kinematically misalignment and the fraction of galaxies with AGN, but did not separate the effect of external interactions and external gas accretion from kinematic misalignment, due to the limited sample size \citep{raimundo23}. With our MaNGA sample we show that even within the sample of galaxies with evidence for interactions, there is an additional increase in the fraction of galaxies with misalignment that host an AGN. The relative difference in AGN fraction in the interacting galaxy sample is similar to what is seen in the full sample (Fig.~\ref{fig:manga_aligned_misaligned}). This shows that while galaxy interactions, such as mergers, may be important for the increase in observed AGN fraction (e.g. \citealt{comerford24}), the dominant mechanism behind the difference in AGN fraction observed in this work (and likely for SAMI - \citealt{raimundo23}), is the presence of kinematically misaligned structures.

Kinematically misaligned structures can only be produced via the external accretion of gas, and therefore galaxy interactions are fundamental to creating the misalignment in the first place. What we show in our work is that for interactions that end up in kinematic misalignment (as opposed to aligned rotation of gas and stars), black holes are more likely in an active phase. This finding indicates that kinematic misalignment plays a major role in the observed fraction of active black holes after an external accretion event.

\section{Conclusions}
In this work we investigate the driving mechanism for the observed higher fraction of AGN residing in galaxies with misaligned gas-to-stellar kinematic axes. We measure a higher fraction of AGN (\totalfracAGNmisaligned) in MaNGA galaxies with strong misalignment between gas and stellar rotation ($\Delta$PA $\ge$ 45$^{\circ}$) than in galaxies with aligned gas and stellar rotation (\totalfracAGNaligned). This result is in line with what was previously found for the smaller sample of galaxies in the SAMI survey \citep{raimundo23}, and shows a connection between the presence of kinematically misaligned gas and a higher fraction of observed black hole activity. 

Kinematically misaligned gas is the consequence of an external accretion event but not all external accretion events result in strongly misaligned stellar to gas kinematics. We therefore investigate for the first time whether the observed increase in the fraction of AGN activity is also driven by the presence of misaligned gas or simply by the external accretion event, irrespective of whether the gas ends up aligned or misaligned. We find that the overall fraction of AGN in interacting galaxies (i.e. those with visual signatures of a past external accretion event) is slightly higher than in the total sample that contains galaxies with and without past interactions. Most importantly, within the sample with evidence for interactions, we still find a higher fraction of AGN in galaxies with misaligned gas (\mergerfracAGNmisaligned for $\Delta$PA $\ge$ 45$^{\circ}$ vs \mergerfracAGNaligned for $\Delta$PA $<$ 45$^{\circ}$). This result indicates that even when comparing galaxies with recent interactions, the misalignment between stellar and gas rotation is associated with an increase of the AGN fraction. The conclusion of this work is that gas-to-stellar kinematic  misalignment is driving the increase in AGN fraction, even within the sample where all galaxies had recent external interactions. This result shows the importance of kinematically misaligned structures to the loss of gas angular momentum and to the fuelling of supermassive black holes.

\section*{Acknowledgments}
We would like to thank the referee for their careful reading of the manuscript and constructive comments. This work was supported by the Science and Technology Facilities Council (STFC) of the UK Research and Innovation via grant reference ST/Y002644/1 and by the European Union's Horizon 2020 research and innovation programme under the Marie Sklodowska-Curie grant agreement No 891744 (SIR). RR thanks Conselho Nacional de Desenvolvimento Cient\'{i}fico e Tecnol\'ogico  (CNPq, Proj. 311223/2020-6,  304927/2017-1 and  400352/2016-8), Funda\c{c}\~ao de amparo \`{a} pesquisa do Rio Grande do Sul (FAPERGS, Proj. 16/2551-0000251-7 and 19/1750-2), Coordena\c{c}\~ao de Aperfei\c{c}oamento de Pessoal de N\'{i}vel Superior (CAPES, Proj. 0001). RAR acknowledges the support from CNPq (Proj. 303450/2022-3, 403398/2023-1, \& 441722/2023-7),  FAPERGS (Proj. 21/2551-0002018-0), and CAPES (Proj. 88887.894973/2023-00). 
M.V. gratefully acknowledges financial support from the Independent Research Fund Denmark via grant number DFF 8021-00130. CRA acknowledges support from the Agencia Estatal de Investigaci\'on of the Ministerio de Ciencia, Innovaci\'on y Universidades (MCIU/AEI) under the grant ``Tracking active galactic nuclei feedback from parsec to kiloparsec scales'', with reference PID2022-141105NB-I00 and the European Regional Development Fund (ERDF).

This research has made use of MaNGA data from SDSS IV. Funding for the Sloan Digital Sky Survey IV has been provided by the Alfred P. Sloan Foundation, the U.S. Department of Energy Office of Science, and the Participating Institutions. SDSS acknowledges support and resources from the Center for High-Performance Computing at the University of Utah. The SDSS web site is www.sdss4.org.
This research has made use of the VizieR catalogue access tool, CDS, Strasbourg, France (DOI : 10.26093/cds/vizier), \cite{ochsenbein00}.
This research made use of Astropy, \href{http://www.astropy.org}{http://www.astropy.org} a community-developed core Python package for Astronomy \cite{astropy:2013}.

\section*{Data Availability}
 
The MaNGA MEGACUBES used in this work are publicly available through a web interface at https://manga.linea.org.br/ and https://manga.if.ufrgs.br/. 



\bibliographystyle{mnras}
\bibliography{AGN}

\bsp	
\label{lastpage}
\end{document}